\documentstyle[twoside, epsfig]{article}

\catcode`\@=11
\long\def\@makefntext#1{
\protect\noindent \hbox to 3.2pt {\hskip-.9pt
$^{{\eightrm\@thefnmark}}$\hfil}#1\hfill}       

\def\@makefnmark{\hbox to 0pt{$^{\@thefnmark}$\hss}}    

\def\ps@myheadings{\let\@mkboth\@gobbletwo
\def\@oddhead{\hbox{}
\rightmark\hfil\eightrm\thepage}
\def\@oddfoot{}\def\@evenhead{\eightrm\thepage\hfil
\leftmark\hbox{}}\def\@evenfoot{}
\def\sectionmark##1{}\def\subsectionmark##1{}}

\oddsidemargin=\evensidemargin
\newcounter{sectionc}\newcounter{subsectionc}\newcounter{subsubsectionc}
\renewcommand{\section}[1] {\vspace{12pt}\addtocounter{sectionc}{1}
\setcounter{subsectionc}{0}\setcounter{subsubsectionc}{0}\noindent
    {\tenbf\thesectionc. #1}\par\vspace{5pt}}
\renewcommand{\subsection}[1] {\vspace{12pt}\addtocounter{subsectionc}{1}
    \setcounter{subsubsectionc}{0}\noindent
    {\bf\thesectionc.\thesubsectionc. {\kern1pt \bfit #1}}\par\vspace{5pt}}
\renewcommand{\subsubsection}[1] {\vspace{12pt}\addtocounter{subsubsectionc}{1}
    \noindent{\tenrm\thesectionc.\thesubsectionc.\thesubsubsectionc.
    {\kern1pt \tenit #1}}\par\vspace{5pt}}
\newcommand{\nonumsection}[1] {\vspace{12pt}\noindent{\tenbf #1}
    \par\vspace{5pt}}

\newcounter{appendixc}
\newcounter{subappendixc}[appendixc]
\newcounter{subsubappendixc}[subappendixc]
\renewcommand{\thesubappendixc}{\Alph{appendixc}.\arabic{subappendixc}}
\renewcommand{\thesubsubappendixc}
    {\Alph{appendixc}.\arabic{subappendixc}.\arabic{subsubappendixc}}

\renewcommand{\appendix}[1] {\vspace{12pt}
        \refstepcounter{appendixc}
        \setcounter{figure}{0}
        \setcounter{table}{0}
        \setcounter{lemma}{0}
        \setcounter{theorem}{0}
        \setcounter{corollary}{0}
        \setcounter{definition}{0}
        \setcounter{equation}{0}
        \renewcommand{\thefigure}{\Alph{appendixc}.\arabic{figure}}
        \renewcommand{\thetable}{\Alph{appendixc}.\arabic{table}}
        \renewcommand{\theappendixc}{\Alph{appendixc}}
        \renewcommand{\thelemma}{\Alph{appendixc}.\arabic{lemma}}
        \renewcommand{\thetheorem}{\Alph{appendixc}.\arabic{theorem}}
        \renewcommand{\thedefinition}{\Alph{appendixc}.\arabic{definition}}
        \renewcommand{\thecorollary}{\Alph{appendixc}.\arabic{corollary}}
        \renewcommand{\theequation}{\Alph{appendixc}.\arabic{equation}}
        \noindent{\tenbf Appendix \theappendixc #1}\par\vspace{5pt}}
\newcommand{\subappendix}[1] {\vspace{12pt}
        \refstepcounter{subappendixc}
        \noindent{\bf Appendix \thesubappendixc. {\kern1pt \bfit #1}}
    \par\vspace{5pt}}
\newcommand{\subsubappendix}[1] {\vspace{12pt}
        \refstepcounter{subsubappendixc}
        \noindent{\rm Appendix \thesubsubappendixc. {\kern1pt \tenit #1}}
    \par\vspace{5pt}}

\newcommand{\textlineskip}{\baselineskip=13pt}
\newcommand{\smalllineskip}{\baselineskip=10pt}

\def\eightcirc{
\begin{picture}(0,0)
\put(4.4,1.8){\circle{6.5}}
\end{picture}}
\def\eightcopyright{\eightcirc\kern2.7pt\hbox{\eightrm c}}

\newcommand{\copyrightheading}[1]
    {\vspace*{-2.5cm}\smalllineskip{\flushleft
    {\footnotesize International Journal of Modern Physics C, #1}\\
    {\footnotesize $\eightcopyright$\, World Scientific Publishing
     Company}\\
     }}

\newcommand{\publisher}[2]{{\begin{center}\footnotesize\smalllineskip
    Received #1
    \end{center}
    }}

\def\abstracts#1#2#3{{
    \centering{\begin{minipage}{4.5in}\baselineskip=10pt\footnotesize
    \parindent=0pt #1\par
    \parindent=15pt #2\par
    \parindent=15pt #3
    \end{minipage}}\par}}

\def\keywords#1{{
    \centering{\begin{minipage}{4.5in}\baselineskip=10pt\footnotesize
    {\footnotesize\it Keywords}\/: #1
    \end{minipage}}\par}}

\renewenvironment{thebibliography}[1]
        {\frenchspacing
     \ninerm\baselineskip=11pt
         \begin{list}{\arabic{enumi}.}
        {\usecounter{enumi}\setlength{\parsep}{0pt}
     \setlength{\leftmargin 12.7pt}{\rightmargin 0pt} 
         \setlength{\itemsep}{0pt} \settowidth
    {\labelwidth}{#1.}\sloppy}}{\end{list}}

\newcounter{itemlistc}
\newcounter{romanlistc}
\newcounter{alphlistc}
\newcounter{arabiclistc}

\newcommand{\fcaption}[1]{
        \refstepcounter{figure}
        \setbox\@tempboxa = \hbox{\footnotesize Fig.~\thefigure. #1}
        \ifdim \wd\@tempboxa > 5in
           {\begin{center}
        \parbox{5in}{\footnotesize\smalllineskip Fig.~\thefigure. #1}
            \end{center}}
        \else
             {\begin{center}
             {\footnotesize Fig.~\thefigure. #1}
              \end{center}}
        \fi}

\newcommand{\tcaption}[1]{
        \refstepcounter{table}
        \setbox\@tempboxa = \hbox{\footnotesize Table~\thetable. #1}
        \ifdim \wd\@tempboxa > 5in
           {\begin{center}
        \parbox{5in}{\footnotesize\smalllineskip Table~\thetable. #1}
            \end{center}}
        \else
             {\begin{center}
             {\footnotesize Table~\thetable. #1}
              \end{center}}
        \fi}

\def\@citex[#1]#2{\if@filesw\immediate\write\@auxout
    {\string\citation{#2}}\fi
\def\@citea{}\@cite{\@for\@citeb:=#2\do
    {\@citea\def\@citea{,}\@ifundefined
    {b@\@citeb}{{\bf ?}\@warning
    {Citation `\@citeb' on page \thepage \space undefined}}
    {\csname b@\@citeb\endcsname}}}{#1}}

\newif\if@cghi
\def\cite{\@cghitrue\@ifnextchar [{\@tempswatrue
    \@citex}{\@tempswafalse\@citex[]}}
\def\citelow{\@cghifalse\@ifnextchar [{\@tempswatrue
    \@citex}{\@tempswafalse\@citex[]}}
\def\@cite#1#2{{$\null^{#1}$\if@tempswa\typeout
    {IJCGA warning: optional citation argument
    ignored: `#2'} \fi}}

\def\pmb#1{\setbox0=\hbox{#1}
    \kern-.025em\copy0\kern-\wd0
    \kern.05em\copy0\kern-\wd0
    \kern-.025em\raise.0433em\box0}

\def\fnt#1#2{\footnotetext{\kern-.3em
    {$^{\mbox{\scriptsize #1}}$}{#2}}}

\def\fpage#1{\begingroup
\voffset=.3in
\thispagestyle{empty}\begin{table}[b]\centerline{\footnotesize #1}
    \end{table}\endgroup}

\def\runninghead#1#2{\pagestyle{myheadings}
\markboth{{\protect\footnotesize\it{\quad #1}}\hfill}
{\hfill{\protect\footnotesize\it{#2\quad}}}}
\font\tenrm=cmr10
\font\tenit=cmti10
\font\tenbf=cmbx10
\font\bfit=cmbxti10 at 10pt
\font\ninerm=cmr9

\font\eightrm=cmr8

\def\qed{\hbox{${\vcenter{\vbox{            
   \hrule height 0.4pt\hbox{\vrule width 0.4pt height 6pt
   \kern5pt\vrule width 0.4pt}\hrule height 0.4pt}}}$}}

\def\bsc{{\sc a\kern-6.4pt\sc a\kern-6.4pt\sc a}}   
\def\bflatex{\bf L\kern-.30em\raise.3ex\hbox{\bsc}\kern-.14em
T\kern-.1667em\lower.7ex\hbox{E}\kern-.125em X}

\begin{document}

\runninghead{Realistic modeling of strongly correlated electron
systems:}{An introduction to the LDA+DMFT approach}
\normalsize\textlineskip \thispagestyle{empty}
\setcounter{page}{1}

\copyrightheading{}         

\vspace*{0.88truein}

\fpage{1} \centerline{\bf} \vspace*{0.035truein} \centerline{\bf
REALISTIC MODELING OF STRONGLY  CORRELATED}
\vspace*{0.035truein} \centerline {\bf ELECTRON SYSTEMS:}
\vspace*{0.035truein} \centerline {\bf AN INTRODUCTION TO THE
LDA+DMFT APPROACH} \vspace*{0.37truein} \centerline{\footnotesize
K. Held$^{1,3}$, I.A.~Nekrasov$^{2}$, N. Bl{\"{u}}mer$^{1}$,
V.I.~Anisimov$^{2}$, and D. Vollhardt$^{1}$} \vspace*{0.15truein}
\centerline{\footnotesize\it $^1$Theoretical Physics III, Center
for Electronic Correlations and Magnetism} \baselineskip=10pt
\centerline{\footnotesize\it Institute for Physics,
University of Augsburg, D-86135 Augsburg, Germany}
\baselineskip=15pt \centerline{\footnotesize\it $^{2}$Institute of
Metal Physics, Russian Academy of Sciences-Ural Division}
\baselineskip=10pt \centerline{\footnotesize\it 620219
Yekaterinburg GSP-170, Russia} \baselineskip=20pt
\centerline{\footnotesize\it $^3$Present address: Physics
Department, Princeton University} \baselineskip=10pt
\centerline{\footnotesize\it Princeton, NJ 08544, USA}
\vspace*{0.225truein}
\publisher{(received date)}

\vspace*{0.21truein} \abstracts{The LDA+DMFT approach merges
conventional band structure theory in the local density
approximation (LDA) with a state-of-the-art many-body technique,
the dynamical mean-field theory (DMFT). This new computational
scheme has recently become a powerful tool for {\em ab initio}
investigations of real materials with strong electronic
correlations. In this paper an introduction to the basic ideas and
the set-up of the LDA+DMFT approach is
given. Results for the photoemission spectra of the transition metal oxide La%
$_{1-x}$Sr$_{x}$TiO$_{3}$, obtained by solving the DMFT-equations
by quantum Monte-Carlo (QMC)\ simulations, are presented and are
found to be in very good agreement with experiment. The
numerically exact DMFT(QMC) solution is compared with results
obtained by two approximative solutions, i.e., the iterative
perturbation theory and the non-crossing approximation. }{}{}

\vspace*{10pt}
\keywords{{71.27.+a}{ Strongly correlated electron systems; heavy fermions};
{74.25.Jb}{ Electronic structure};  {79.60.-i} { Photoemission and photoelectron spectra}}


\vspace*{1pt}\textlineskip  

\vspace*{1pt}\textlineskip
\section{Introduction}
\vspace*{-0.5pt}
\noindent

It is an established fact of solid state physics that many
electronic properties of matter are well described by the purely
electronic Hamiltonian
\begin{eqnarray}
\hat{H} &=&\sum_{\sigma }\int \!d^{3}r\;\hat{\Psi}^{+}({\bf
r},\sigma )\left[
{-\frac{\hbar ^{2}}{2m_{e}}\Delta +V_{{\rm ion}}({\bf r})}\right] \hat{\Psi}(%
{\bf r},\sigma )  \nonumber \\
&&+\frac{1}{2}\sum_{\sigma \sigma ^{\prime }}\int \!d^{3}r \, d^{3}r^{\prime }\;
\hat{\Psi}^{+}({\bf r},\sigma )\hat{\Psi}^{+}({\bf r^{\prime
}},\sigma
^{\prime })\;{V_{{\rm ee}}({\bf r}\!-\!{\bf r^{\prime }})}\;\hat{\Psi}({\bf %
r^{\prime }},\sigma ^{\prime })\hat{\Psi}({\bf r},\sigma ),
\label{abinitioHam}
\end{eqnarray}%
where the crystal lattice enters only through an ionic potential.
The applicability of this approach may be justified by the
validity of the Born and Oppenheimer approximation\cite{Born27a}.
Here, $ \hat{\Psi}^{+}({\bf r},\sigma )$ and $\hat{\Psi}({\bf
r},\sigma )$ are field operators that create and annihilate an
electron at position ${\bf r}$ with spin $\sigma $, $\Delta $ is
the Laplace operator, $m_{e}$ the electron mass, $e$ the electron
charge, and
\begin{eqnarray}
V_{{\rm ion}} ({\rm r})=-e^2\sum_i \frac{Z_i}{|{\bf r}-{\bf R_i}|}
& {\rm and} & V_{\rm ee}({\bf r}\!-\!{\bf r'})=\frac{e^2}{2}
\sum_{{\bf r} \neq {\bf r'}} \frac{1}{|{\bf r}-{\bf r'}|}
\end{eqnarray}
denote the one-particle potential due to all ions $i$
with charge $eZ_{i}$ at given positions ${\bf R_{i}}$, and the
electron-electron interaction, respectively.

While the {\em ab initio }Hamiltonian (\ref{abinitioHam}) is easy
to write down it is impossible to solve, even numerically, if more
than a few electrons are involved. The reason for this is the
electron-electron interaction which correlates every electron with
all others. Therefore, one either needs to make substantial
approximations to deal with the Hamiltonian (\ref{abinitioHam}),
or replace it by a strongly simplified model Hamiltonian. At
present these different strategies for the investigation of the
electronic properties of solids are applied by two largely
separate groups: the density functional theory (DFT) and the
many-body community. It is known for a long time already that DFT,
together with its local density approximation (LDA), is a highly
successful technique for the calculation of the electronic
structure of many real materials\cite{JonesGunn}. However, for
strongly correlated materials, i.e., $d$- and $f$-electron system
which have a Coulomb interaction comparable to the band-width,
DFT/LDA is seriously restricted in its accuracy and reliability.
Here, the model Hamiltonian approach is more general and powerful
since there exist systematic theoretical techniques to investigate
the many-electron problem with increasing accuracy. These
many-body techniques allow to describe qualitative tendencies and
understand the basic mechanism of various physical phenomena. At
the same time the model Hamiltonian approach is seriously
restricted in its ability to make quantitative predictions since
the input parameters are not accurately known and hence need to be
adjusted. One of the most successful techniques in this respect is
the dynamical mean-field theory (DMFT) -- a non-perturbative
approach to strongly
correlated electron systems which was developed during the past decade\cite%
{MetzVoll89,vollha93,pruschke,georges96}. The LDA+DMFT approach,
which was first formulated by Anisimov et al.\cite{poter97},
combines the strength of DFT/LDA to describe the weakly correlated
part of the {\em ab initio} Hamiltonian (\ref{abinitioHam}), i.e.,
electrons in $s$- and $p$-orbitals as well as the long-range
interaction of the $d$- and $f$-electrons, with the power of DMFT
to describe the strong correlations induced by the local Coulomb
interaction of the $d$- or $f$-electrons.

Starting from the {\em ab initio} Hamiltonian (\ref{abinitioHam}),
the LDA+DMFT approach is derived in section 2. As a particular
example, the LDA+DMFT calculation for La$_{1-x}$Sr$_{x}$TiO$_{3}$
is discussed in section 3. Furthermore, the LDA+DMFT spectrum is
calculated by means of numerically exact quantum Monte Carlo (QMC)
simulations, and is compared to that obtained within two
approximations commonly employed to solve the DMFT, i.e., the
iterated perturbation theory (IPT) and the non-crossing
approximation (NCA). A discussion of the LDA+DMFT approach and its
future prospects (section 4) closes the presentation.

\section{The LDA+DMFT approach}

\noindent \label{LDADMFT} \vspace*{-9.5pt}

\subsection{Density functional theory}

\noindent
The fundamental theorem of DFT by Hohenberg and
Kohn\cite{Hohenberg64a} states that the ground state energy is a
functional of the electron density which assumes its minimum at
the ground state electron density.
 Following
Levy\cite{Levy}, this theorem is easily proved and the functional
even constructed by taking the minimum (infimum) of the energy
expectation value w.r.t. all (many-body) wave functions $\varphi
({\bf r_{1}}\sigma _{1},...{\bf r_{N}}\sigma _{N})$ at a given
electron number $N$ which yield the electron density $\rho ({\bf
r})$:
\begin{equation}
E[\rho ]=\inf \Big\{\langle \varphi |\hat{H}|\varphi \rangle \;\;\Big|%
\;\;\langle \varphi |\sum_{i=1}^{N}\delta ({\bf r}-{\bf
r_{i}})|\varphi \rangle =\rho ({\bf r})\Big\}.
\end{equation}%
However, this construction is of no practical value since it
actually requires the evaluation of the Hamiltonian
(\ref{abinitioHam}). Only certain
contributions like the Hartree energy $E_{{\rm Hartree}}[\rho ]=\frac{1}{2}%
\int d^{3}r^{\prime }\,d^{3}r\;V_{{\rm ee}}({\bf r}\!-\!{\bf r^{\prime }}%
)\;\rho ({\bf r^{\prime })\rho (r)}$ and the energy of the ionic potential $%
E_{{\rm ion}}[\rho ]=\int d^{3}r\;V_{{\rm ion}}({\bf r})\;\rho
({\bf r)}$ can be expressed directly in terms of the electron
density. This leads to
\begin{equation}
E[\rho ]=E_{{\rm kin}}[\rho ]+E_{{\rm ion}}[\rho ]+E_{{\rm
Hartree}}[\rho ]+E_{{\rm xc}}[\rho ],  \label{Erho}
\end{equation}%
where $E_{{\rm kin}}[\rho ]$ denotes the kinetic energy, and $E_{{\rm xc}%
}[\rho ]$ is the unknown exchange and correlation term which
contains the energy of the electron-electron interaction except
for the Hartree term. Hence all the difficulties of the many-body
problem have been transfered into $E_{{\rm xc}}[\rho ]$. While the
kinetic energy $E_{{\rm kin}}$ cannot be expressed explicitly in
terms of the electron density one can employ a trick to determine
it. Instead of minimizing $E[\rho ]$ w.r.t. $\rho $ one minimizes
it w.r.t. a set of one-particle wave functions $\varphi _{i}$
related to $\rho $ via
\begin{equation}
\rho ({\bf r})=\sum_{i=1}^{N}|\varphi _{i}({\bf r})|^{2}.
\label{rhophi}
\end{equation}%
To guarantee the normalization of $\varphi _{i}$, the Lagrange parameters $%
\varepsilon _{i}$ are introduced such that the variation $\delta
\{E[\rho ]+\varepsilon _{i}[1-\int d^{3}r|\varphi _{i}({\bf
r})|^{2}]\}/\delta \varphi _{i}({\bf r})=0$ yields the
Kohn-Sham\cite{Kohn65} equations:
\begin{equation}
\left[ -{\frac{\hbar ^{2}}{2m_{e}}\Delta +V_{{\rm ion}}({\bf r})}+\int d^{3}{r^{\prime }}\,{\rho ({\bf r^{\prime }})}{V_{{\rm ee}}({\bf r}\!-\!{\bf %
r^{\prime }})}+{{\frac{\delta {E_{{\rm xc}}[\rho ]}}{\delta \rho ({\bf r)}}}}%
\right] \varphi _{i}({\bf r})=\varepsilon _{i}\;\varphi _{i}({\bf
r}). \label{KohnSham}
\end{equation}%
These equations have the same form as a one-particle
Schr\"{o}dinger equation which, {\em a posteriori}, justifies to
calculate the kinetic energy by means of the one-particle
wave-function ansatz. The kinetic energy
at the ground state density is, then, given by $E_{{\rm kin}}[\rho _{{\rm min%
}}]=-\sum_{i=1}^{N}\langle \varphi _{i}|{\hbar ^{2}\Delta }/{(2m_{e})}%
|\varphi _{i}\rangle $ if the $\varphi _{i}$ are the
self-consistent (spin-degenerate) solutions of Eqs.
(\ref{KohnSham}) and (\ref{rhophi}) with lowest ``energy''
$\epsilon _{i}$. Note, however, that the one-particle potential of
Eq. (\ref{KohnSham}), i.e.,
\begin{equation}
V_{{\rm ion}}({\bf r})+\int d^{3}{\ r^{\prime }}{\rho ({\bf r^{\prime }})}{%
V_{{\rm ee}}({\bf r}\!-\!{\bf r^{\prime }})}+{{\frac{\delta {E_{{\rm xc}%
}[\rho ]}}{\delta \rho ({\bf r)}}}},
\end{equation}%
is only an auxiliary potential which artificially arises in the
approach to minimize $E[\rho ]$. Thus, the wave functions $\varphi
_{i}$ and the Lagrange parameters $\varepsilon _{i}$ have no
physical meaning at this point.

\subsection{Local density approximation}

\noindent So far no approximations have been employed since the
difficulty
of the many-body problem was only transferred to the unknown functional $%
E_{xc}[\rho ]$. For this term the local-density approximation
(LDA) which approximates the functional $E_{xc}[\rho ]$ by a
function that depends on the local density only, i.e.,
\begin{equation}
E_{xc}[\rho ]\;{\rightarrow }\;\int d^{3}r\;E_{xc}^{{\rm LDA}}(\rho ({\bf r}%
)),
\end{equation}%
was found to be unexpectedly successful. Here, $E_{xc}^{{\rm LDA}}(\rho (%
{\bf r}))$ is usually calculated from the Hartree-Fock solution or the
numerical
simulation of the jellium problem which is defined by $V_{{\rm ion}}({\bf r})=%
{\rm const}$\cite{jellium}.

In principle DFT/LDA only allows one to calculate static
properties like the ground state energy or its derivatives.
However, one of the major applications of LDA is the calculation
of band structures. To this end, the Lagrange parameters
$\varepsilon _{i}$ {\em are interpreted} as the physical
(one-particle) energies of the system under consideration. Since
the true ground-state is not a simple one-particle wave-function,
this is a further approximation beyond DFT. Actually, this
approximation corresponds to the replacement of the Hamiltonian (\ref%
{abinitioHam}) by
\begin{eqnarray}
\hat{H}_{{\rm LDA}} &=&\sum_{\sigma }\int \!d^{3}r\;\hat{\Psi}^{+}({\bf r}%
,\sigma )\left[ -\frac{\hbar ^{2}}{2m_{e}}\Delta +V_{{\rm ion}}({\bf r}%
)+\int d^3{r^{\prime }}\,{\rho ({\bf r^{\prime }})}{V_{{\rm ee}}({\bf r}\!-\!%
{\bf r^{\prime }})}\right.   \nonumber \\ &&\left. \phantom
{\sum_{\sigma \sigma'} \int\! d^3r\; \hat{\Psi}^+({\bf r},\sigma)
\big[ \;}+{{\frac{\delta {E_{xc}^{{\rm LDA}}}[\rho ]}{\delta \rho
({\bf r)}}}}\right] \hat{\Psi}({\bf r},\sigma ).  \label{HLDA0}
\end{eqnarray}%
For practical calculations one needs to expand the field operators
w.r.t. a basis $\Phi _{ilm}$, e.g., a linearized muffin-tin
orbital (LMTO)\cite{LMTO} basis (here $i$ denotes lattice sites;
$l$ and $m$ are orbital indices). In this basis,
\begin{eqnarray}
\hat{\Psi}^+({\bf r},\sigma) &=& \sum_{i l m} \hat{c}_{i l
m}^{\sigma \dagger} \Phi_{i l m}({\bf r})^{\phantom{+}}
\end{eqnarray}
such that the Hamiltonian (\ref{HLDA0}) reads
\begin{eqnarray}
 \hat{H}_{\rm LDA}  &=& \sum_{ilm,{\rm }jl^{\prime }m^{\prime },\sigma }(\delta
_{ilm,jl^{\prime }m^{\prime }} \;
{\varepsilon_{ilm}}^{\phantom{\sigma}}{\hat{n}}_{ilm}^{\sigma
}+ {t_{ilm,jl^{\prime }m^{\prime }}}\;{\hat{c}}_{ilm}^{\sigma \dagger }{\hat{c}}%
_{jl^{\prime }m^{\prime }}^{\sigma }). \label{HLDA}
\end{eqnarray}

Here, ${\hat{n}}_{ilm}^{\sigma }=\hat{c}_{ilm}^{\sigma \dagger }{\hat{c}}%
_{ilm}^{\sigma }$,
\begin{equation}
t_{ilm,jl^{\prime }m^{\prime }}=\Big\langle\Phi
_{ilm}\Big|-\frac{\hbar
^{2}\Delta }{2m_{e}}+V_{{\rm ion}}({\bf r})+\int d{\ r^{\prime }}{\rho ({\bf %
r^{\prime }})}{V_{{\rm ee}}({\bf r}\!-\!{\bf r^{\prime }})}+{{\frac{\delta {%
E_{xc}^{{\rm LDA}}}[\rho ]}{\delta \rho ({\bf r)}}}}\Big|\Phi
_{jl^{\prime }m^{\prime }}\Big\rangle
\end{equation}%
for $ilm\neq jl^{\prime }m^{\prime }$ and zero otherwise; $\varepsilon _{ilm}$
denotes the corresponding diagonal part.

As for static properties LDA band structure calculations are also
often
highly successful -- but only for weakly correlated materials\cite{JonesGunn}%
. Indeed, the self-consistent solution of the one-particle Hamiltonian $%
\hat{H}_{{\rm LDA}}$ (\ref{HLDA}) together with Eq.~(\ref{rhophi})
treats electronic {\em correlations} only rudimentarily.
Consequently, LDA is not reliable when applied to correlated
materials, and can even be completely wrong. For example, it
predicts the antiferromagnetic insulator La$_{2}$CuO$_{4}$ to be a
non-magnetic metal\cite{LDAfailure}.

\subsection{Supplementing LDA with local Coulomb correlations}

\noindent Of prime importance for correlated materials are the
local Coulomb interactions between $d$- and $f$-electrons on the
same lattice site since these contributions are largest. This is
due to the extensive overlap of
these localized orbitals which results in strong correlations\cite{nonlocalU}%
. To correct for these contributions, one can supplement the LDA
Hamiltonian (\ref{HLDA}) with the local Coulomb interaction
$U_{mm^{\prime }}^{\sigma
\sigma ^{\prime }}$ between the localized electrons (for which we assume $%
i=i_{d}$ and $l=l_{d}$):
\begin{eqnarray} \hat{H}_{\rm LDA+correl} &=&
\hat{H}_{\rm LDA} +
 \frac{1}{2} {\sum_{i=i_d, l= l_d,m\sigma m^{\prime }\sigma
}}^{\!\!\!\!\!\!\!\!\!\!\!\!\!\!\!\prime } \;\; {U_{mm^{\prime
}}^{\sigma \sigma'}} \hat{n}_{ilm\sigma } \hat{n}_{ilm^{\prime
}\sigma^{\prime }} -  \hat{H}^{U}_{\rm LDA}. \label{Hint}
\end{eqnarray}
 Here, the prime on the sum indicates that
at least two of the indices of an operator have to be different,
and a term $\hat{H}_{{\rm LDA}}^{U}$ is substracted to avoid
double-counting of those contributions of the local Coulomb
interaction already contained in $\hat{H}_{{\rm LDA}}$. Since
there does not exist a direct microscopic or diagrammatic link
between the model
Hamiltonian approach and LDA it is not possible to express $\hat{H}_{{\rm LDA%
}}^{U}$ rigorously in terms of $U$ and $\rho $. Guided by the
observation that the LDA calculates the {\em total energy} of
isolated atoms rather well, it was argued\cite{Anisimov91} that
the average energy $E_{{\rm LDA}}^{U}$ corresponding to
$\hat{H}_{{\rm LDA}}^{U}$ is well approximated by the energy of
the interaction term in the atomic limit. Hence, in the case of an
orbital- and spin-independent $U_{mm^{\prime }}^{\sigma \sigma
^{\prime }}=U$ one may write
\begin{equation}
E_{{\rm LDA}}^{U}=\frac{1}{2}{U}n_{d}(n_{d}-1).  \label{ELDAU}
\end{equation}%
(For the corresponding equation including Hund's rule coupling see Ref. \cite{poter97}). Here, $n_{d}=\sum_{m}n_{il_{d}m}=\sum_{m}\langle \hat{n}%
_{il=l_{d}m}\rangle $ is the total number of interacting
electrons. Since the one-electron LDA energies can be obtained
from the derivatives of the total energy w.r.t. the occupation
numbers of the corresponding states, the
one-electron energy level for the {\em non-interacting} states of (\ref{Hint}%
) is obtained as\cite{Anisimov91}
\begin{equation}
\varepsilon _{il_{d}m}^{0}:=\frac{{\rm d}}{{\rm d}n_{il_{d}m}}(E_{{\rm LDA}%
}-E_{{\rm LDA}}^{U})=\varepsilon _{il_{d}m}-U(n_{d}-\frac{1}{2})
\label{neweps}
\end{equation}%
where $\varepsilon _{il_{d}m}$ is defined in (\ref{HLDA}) and
$E_{{\rm LDA}}$ is the total energy calculated from $\hat{H}_{{\rm
LDA}}$ (\ref{HLDA}).

This leads to a new Hamiltonian describing the non-interacting system%
\begin{eqnarray}
H^{0}_{{\rm LDA}}&=&\sum_{ilm,jl^{\prime }m^{\prime },\sigma
}\!\!\!(\delta _{ilm,jl^{\prime }m^{\prime }}\varepsilon
_{ilm}^{0}{\hat{n}}_{ilm}^{\sigma
}+t_{ilm,jl^{\prime }m^{\prime }}{\hat{c}}_{ilm}^{\sigma \dagger }{\hat{c}}%
_{jl^{\prime }m^{\prime }}^{\sigma }),
 \end{eqnarray}
where $\varepsilon _{il_{d}m}^{0}$ is given by (\ref{neweps}) for
the interacting orbitals and $\varepsilon _{ilm}^{0}=\varepsilon
_{ilm}$ for the non-interacting orbitals. 
While it is not clear at  present how to systematically
subtract $\hat{H}_{{\rm LDA}}^{U}$ one should note that the
 subtraction of a Hartree-type energy does not substantially 
affect the {\em overall}
behavior of a strongly correlated paramagnetic metal in the
vicinity of a Mott-Hubbard metal-insulator transition (see also
Sec. 2.6).

In the following it is convenient to work in reciprocal space
where the matrix elements of $\hat{H}_{{\rm LDA}}^{0}$ are given
by
\begin{eqnarray}
(H^{0}_{{\rm LDA}}({\bf k}))_{qlm,q^{\prime }l^{\prime }m^{\prime
}}\!&\!=\!&\!(H_{{\rm LDA}}({\bf k}))_{qlm,q^{\prime}l^{\prime
}m^{\prime }}
 -\delta _{qlm,q^{\prime }l^{\prime}m^{\prime }}\delta _{ql,q_{d}l_{d}}U(n_{d}-\frac{1}{2}).
\end{eqnarray}
Here, $q$ is an index of the atom in the elementary unit cell, $(H_{{\rm LDA}%
}({\bf k}))_{qlm,q^{\prime }l^{\prime }m^{\prime }}$ is the matrix
element of (\ref{HLDA}) in ${\bf k}$-space, and $q_{d}$ denotes
the atoms with
interacting orbitals in the unit cell. The non-interacting part, $\hat{H}_{%
{\rm LDA}}^{0}$, supplemented with the local Coulomb interaction
forms the (approximated) {\em ab initio} Hamiltonian for a
particular material under investigation:
\begin{equation}
\hat{H}_{{\rm LDA+correl}}=H_{{\rm
LDA}}^{0}+{\sum_{i=i_{d},l=l_{d},m\sigma
m^{\prime }\sigma }}^{\!\!\!\!\!\!\!\!\!\!\!\!\!\!\!\prime }\;\;\;{%
U_{mm^{\prime }}^{\sigma \sigma ^{\prime }}}\, \hat{n}_{ilm\sigma }\hat{n}%
_{ilm^{\prime }\sigma ^{\prime }}.  \label{HLDAcor}
\end{equation}%
To make use of this {\em ab initio }Hamiltonian it is still
necessary to determine the Coulomb interaction $U$. To this end,
one can calculate the LDA ground state energy for different
numbers of interacting elecrons $n_{d}$ (''constrained LDA''\cite{Parameters}) and
employ Eq. (\ref{ELDAU}) whose second derivative w.r.t. $n_{d}$
yields $U$. However, one should keep in mind that, while the total
LDA spectrum is rather insensitive to the choice of the basis, the
calculation of $U$ strongly depends on the shape of the orbitals
which are considered to be interacting. Thus, an appropriate basis
like LMTO is mandatory and, even so, some uncertainty in $U$
remains.

\subsection{Dynamical mean-field theory}

\noindent The many-body extension of LDA, Eq. (\ref{HLDAcor}), was
proposed by Anisimov et al.\cite{Anisimov91} in the context of
their LDA+U approach. Within LDA+U the Coulomb interactions of
(\ref{HLDAcor}) are treated within the Hartree-Fock approximation;
hence it does not contain true many-body physics. While LDA+U is
successful in describing long-range ordered, insulating states of
correlated electronic systems it fails to describe strongly correlated
{\em %
paramagnetic} states. To go beyond LDA+U and capture the many-body
nature of the electron-electron interaction, i.e., the frequency
dependence of the self-energy, various approximation schemes have
been proposed and applied
recently\cite{poter97,lichten98,janis98,laegsgaard,wolenski98,zoelfl99}.
One
of the most promising approaches, first implemented by Anisimov et
al.\cite%
{poter97}, is to solve (\ref{HLDAcor}) within DMFT\cite%
{MetzVoll89,vollha93,pruschke,georges96} (''LDA+DMFT''). Of all
extensions of LDA only the LDA+DMFT approach is presently able to
describe the physics of {\em strongly} correlated, paramagnetic
metals with well-developed upper and lower Hubbard bands and a
narrow quasiparticle peak, which is determined by the vicinity of
a Mott-Hubbard metal-insulator transition.

During the last ten years DMFT has proved to be a successful
approach to
investigate strongly correlated systems with local Coulomb
interactions\cite%
{georges96}. It becomes exact in the limit of high lattice
coordination numbers\cite{MetzVoll89} and preserves the dynamics
of local interactions. Hence, it represents a {\em dynamical}
mean-field approximation. In this non-perturbative approach the
lattice problem is mapped onto an effective single-site problem
which has to be determined self-consistently together
with the ${\bf k}$-integrated Dyson equation connecting the self energy
$%
\Sigma $ and the Green function $G$ at frequency $\omega $:
\begin{eqnarray}
G_{qlm,q^{\prime }l^{\prime }m^{\prime }}(\omega )=\!\frac{1}{V_{B}}\int
{{%
d^{3}}{k}} \!&\!\left[ \;\omega \;\delta _{qlm,q^{\prime }l^{\prime
}m^{\prime }}-(H_{{\rm LDA}}^{0}({\bf k}))_{qlm,q^{\prime
}l^{\prime }m^{\prime }}\right.   &\nonumber \\ &\;\;-\;\delta
_{ql,q_{d}l_{d}}\;\Sigma _{qlm,q^{\prime }l^{\prime }m^{\prime
}}(\omega )]^{-1}.&  \label{Dyson}
\end{eqnarray}%
Here, $[...]^{-1}$ implies the inversion of the matrix with elements $n$
(=$%
qlm$), $n^{\prime }$(=$q^{\prime }l^{\prime }m^{\prime }$), and
integration extends over the Brillouin zone with volume $V_{B}$.

The DMFT single-site problem depends on ${\cal
G}^{-1}=G^{-1}+\Sigma $ and is equivalent to an Anderson impurity
problem\cite{georges92,Jarrell92a} if its hybridization $\Delta
(\omega )$ satisfies ${\cal G}^{-1}(\omega )=\omega -\int d\omega
^{\prime }\Delta (\omega ^{\prime })/(\omega -\omega ^{\prime })$.
The local one-particle Green function at a Matsubara frequency
$\omega _{\nu }=(2\nu +1)\pi /\beta $, orbital index $m$
($l=l_{d}$, $q=q_{d} $), and spin $\sigma $ is given by the
following functional integral over Grassmann variables $\psi
^{\phantom\ast }$ and $\psi ^{\ast }$:
\begin{equation}
G_{\nu m}^{\sigma }=-\frac{1}{{\cal Z}}\int {\cal D}[\psi ]{\cal
D}[\psi
^{\ast }]\psi _{\nu m}^{\sigma \phantom\ast }\psi _{\nu m}^{\sigma \ast
}e^{%
{\cal A}[\psi ,\psi ^{\ast },{\cal G}^{-1}]}.  \label{siam}
\end{equation}%
Here, the single-site action ${\cal A}$ has the form 
\begin{eqnarray}
\lefteqn{{\cal A}[\psi ,\psi ^{\ast },{\cal G}^{-1}]=\sum_{\nu
,\sigma ,m}\psi _{\nu
m}^{\sigma \ast }({\cal G}_{\nu m}^{\sigma })^{-1}\psi _{\nu m}^{\sigma
{%
\phantom\ast }}}\nonumber \\ &&-\frac{1}{2}\!\!\!{\sum_{m\sigma
,m\sigma ^{\prime}}}^{\!\!\!\!\!'}\;\; U_{m m'}^{\sigma \sigma'}
    \int\limits_{0}^{\beta }d\tau \,\psi _{m}^{\sigma \ast }(\tau )\psi
_{m}^{\sigma \phantom\ast }(\tau )\psi _{m^{\prime }}^{\sigma
^{\prime }\ast }(\tau )\psi _{m^{\prime }}^{\sigma ^{\prime }\phantom%
\ast }(\tau ).
\end{eqnarray}
Due to its equivalence to an Anderson impurity problem a variety
of approximative techniques have been employed to solve the DMFT
equations, such as the iterated perturbation theory (IPT)\cite%
{georges92,georges96} and the non-crossing approximation (NCA) \cite%
{NCA1,pruschke89,NCA2}, as well as numerical techniques like
quantum
Monte-Carlo simulations (QMC)\cite{QMC}, exact diagonalization
(ED)\cite%
{caffarel94,georges96}, or numerical renormalization group (NRG)\cite%
{NRG}. IPT is non-self-consistent
second-order perturbation theory in $U$ for the Anderson impurity
problem (%
\ref{siam}) at half-filling. It represents an ansatz that also
yields the correct perturbational $U^{2}$-term and the correct
atomic limit for the self-energy off half-filling\cite{Kajueter}. NCA is a
resolvent perturbation theory in the hybridization parameter
$\Delta (\omega )$ of the Anderson impurity problem. Thus, it is
reliable if the Coulomb interaction $U$ is large compared to the
band-width. In essence, the QMC technique maps the interacting
electron problem (\ref{siam}) onto a sum of non-interacting
problems by means of Hubbard-Stratonovich transformations and
evaluates this sum by Monte-Carlo sampling\cite{QMCdb}. 
 ED directly diagonalizes the Anderson impurity problem at a limited
number of lattice sites.
NRG first replaces the  conduction
band by a discrete set of states at $D \Lambda^{-n}$
($D$: bandwidth; $n=0,...,{\cal N}$) and then 
diagonalizes this problem  iteratively with increasing accuracy at low
energies, i.e., with increasing ${\cal N}$.

In principle, QMC, ED, and NRG are exact methods, but they require
an extrapolation, i.e., the discretization of the imaginary time
$\Delta \tau \rightarrow 0$ (QMC), the number of lattice sites of
the respective impurity model $n_{s}\rightarrow \infty $ (ED), or
the parameter for logarithmic discretization of the conducting
band $\Lambda \rightarrow 1$ (NRG), respectively.

In the present paper, we will not present further details of these
methods and refer the reader to the
literature: IPT\cite{Kajueter,poter97,lichten98}, NCA\cite{zoelfl99},
and QMC%
\cite{rozenberg,Nekrasov00}. In the context of LDA+DMFT we refer
to the computational schemes to solve the DMFT equations
discussed above as LDA+DMFT(X) where X=IPT\cite{poter97}, NCA\cite%
{zoelfl99}, QMC\cite{Nekrasov00} have been investigated in the
case of La$_{1-x}$Sr$_{x}$TiO$_{3}$. The same strategy was
formulated by Lichtenstein and Katsnelson\cite{lichten98} as one
of their LDA++ approaches. Lichtenstein and Katsnelson applied
LDA+DMFT(IPT)\cite{Kats98}, and were the first to use
LDA+DMFT(QMC)\cite{kats99}, to investigate the spectral properties
of iron. Liebsch and Lichtenstein also applied LDA+DMFT(QMC) to
calculate the photoemission spectrum of
Sr$_{2}$RuO$_{4}$\cite{liebsch00}.

\subsection{Self-consistent LDA+DMFT}

In general, the DMFT solution will result in a change of the
occupation of the different bands involved. This changes the
electron density $\rho ({\bf r})$ and, thus, results in a new
LDA-Hamiltonian $H_{\rm LDA}$ (\ref{HLDA}) since $H_{\rm LDA}$
depends on $\rho ({\bf r})$. At the same time also the Coulomb
interaction $U$ changes and needs to be determined by a new
constrained LDA calculation. 
In  a self-consistent LDA+DMFT scheme, $H_{\rm LDA}$
and $U$
define a new Hamiltonian (\ref{HLDAcor}) which again needs to be
solved within DMFT, etc.,
 until convergence is reached:
\begin{equation}
\setlength{\unitlength}{3947sp}%
\begingroup\makeatletter\ifx\SetFigFont\undefined%
\gdef\SetFigFont#1#2#3#4#5{%
  \reset@font\fontsize{#1}{#2pt}%
  \fontfamily{#3}\fontseries{#4}\fontshape{#5}%
  \selectfont}%
\fi\endgroup%
\begin{picture}(3762,497)(901,70)
\thinlines
\put(4651,164){\line( 0,-1){225}}
\put(4651,-61){\line(-1, 0){2550}}
\put(2101,-61){\vector( 0, 1){225}}
\put(3926,314){\vector( 1, 0){600}}
\put(1001,314){\vector( 1, 0){600}}
\put(2551,314){\vector( 1, 0){900}}
\put(901,239){\makebox(0,0)[lb]{\smash{\SetFigFont{12}{14.4}{\rmdefault}{\mddefault}{\updefault}$\!\!\!\! \!\!\! \!\!\! \rho({\bf r}) \;\;\;$}}}
\put(2701,389){\makebox(0,0)[lb]{\smash{\SetFigFont{12}{14.4}{\rmdefault}{\mddefault}{\updefault}DMFT}}}
\put(1876,239){\makebox(0,0)[lb]{\smash{\SetFigFont{12}{14.4}{\rmdefault}{\mddefault}{\updefault}$\!\!\!\!\!\!H_{\rm LDA}$, $U\;\;\;$ }}}
\put(3526,239){\makebox(0,0)[lb]{\smash{\SetFigFont{12}{14.4}{\rmdefault}{\mddefault}{\updefault}$n_{ilm}\;\;\;$}}}
\put(4576,239){\makebox(0,0)[lb]{\smash{\SetFigFont{12}{14.4}{\rmdefault}{\mddefault}{\updefault}$\rho({\bf r})$}}}
\end{picture}
\vspace{.5em}
\end{equation}
Without Coulomb interaction ($U=0$) this scheme reduces to the
self-consistent solution of the Kohn-Sham equations.

\subsection{Simplifications for transition-metal oxides}

\noindent Many transition metal oxides are cubic perovskites, with
only a slight distortion of the cubic crystal structure. In these
systems the transition metal $d$-orbitals lead to strong Coulomb
interactions between the electrons. The cubic crystal-field of the
oxygen causes the $d$-orbitals
to split into three degenerate $t_{2g}$- and two degenerate $e_{g}$%
-orbitals. This splitting is often so strong that the $t_{2g}$- or $e_{g}$%
-bands at the Fermi energy are rather well separated from all
other bands. In this situation the low-energy physics is well
described by taking only the degenerate 
bands at the Fermi energy into
account. Without symmetry breaking, the
Green function and the self-energy of these bands 
remain degenerate, i.e., $%
G_{qlm,q^{\prime }l^{\prime }m^{\prime }}(z)=G(z)\delta
_{qlm,q^{\prime }l^{\prime }m^{\prime }}$ and $\Sigma
_{qlm,q^{\prime }l^{\prime }m^{\prime }}(z)=\Sigma (z)\delta
_{qlm,q^{\prime }l^{\prime }m^{\prime }}$ for $l=l_{d} $ and
$q=q_{d}$ (where $l_{d}$ and $q_{d}$ denote the electrons in the
interacting band at the Fermi energy). Downfolding to a basis with
these
degenerate $q_{d}$-$l_{d}$-bands results in an effective Hamiltonian $H_{%
{\rm LDA}}^{0\;{\rm eff}}$ (where indices $l=l_{d}$ and $q=q_{d}$
are suppressed) \begin{eqnarray} G_{m m'}(\omega)
&=&\frac{1}{{V_{B}}} \int {\rm d}^3 k \;
[(\omega\!-\!\Sigma(\omega))\delta_{m,m^{\prime}}\!-\!(H^{0 \; \rm
eff}_{{\rm LDA}}({\bf k}))_{m,m^{\prime }} ]^{-1}.
\end{eqnarray}
 Due to the diagonal structure of the self-energy
the degenerate interacting Green function can be expressed via the
 non-interacting Green function $G^{0}(\omega)$:
\begin{eqnarray}
G(\omega)&\!=\!&G^{0}(\omega-\Sigma (\omega))=\int d\epsilon
\frac{N^{0}(\epsilon )}{\omega-\Sigma (\omega)-\epsilon}.
\label{intg}
\end{eqnarray}
Thus, it is possible to use the Hilbert transformation of the
unperturbed
LDA-calculated density of states (DOS) $N^{0}(\epsilon )$, i.e., Eq. (\ref%
{intg}), instead of Eq. (\ref{Dyson}). This simplifies the
calculations considerably. With Eq. (\ref{intg}) also some
conceptional simplifications arise: (i) the substraction of
$\hat{H}_{{\rm LDA}}^{U}$ in (\ref{intg}) only results in an
(unimportant) shift of the chemical potential and, thus, the exact
form of $\hat{H}_{{\rm LDA}}^{U}$ is irrelevant; (ii) Luttinger's
theorem of Fermi pinning holds, i.e., the interacting DOS at the
Fermi energy is fixed at the value of the non-interacting DOS at
$T=0$ within a Fermi liquid; (iii) as the number of electrons
within the different bands is fixed, the LDA+DMFT approach is
automatically self-consistent.

In this context it should be noted that the approximation Eq.
(\ref{intg}) is justified only if the overlap between the $t_{2g}$
orbitals and the other orbitals is rather weak.

\section{An example: La$_{1-x}$Sr$_{x}$TiO$_{3}$}

\noindent \label{LaTiO3} The stoichiometric compound LaTiO$_{3}$
is a cubic perovskite with a small orthorhombic distortion
($\angle ~Ti-O-Ti~\approx
~155^{\circ }$)\cite{maclean79} and is an antiferromagnetic insulator\cite%
{eitel86} below $T_{N}=125$~K\cite{gopel}. Above $T_{N}$, or at
low Sr-doping $\ x$, and neglecting the small orthorhombic
distortion (i.e. considering a cubic structure with the same
volume), LaTiO$_{3}$ is a
strongly correlated, but otherwise simple paramagnet with only one 3$d$%
-electron on the trivalent Ti sites. This makes the system a
perfect trial candidate for the LDA+DMFT approach.

\begin{figure}[tbp]
 \centering
\includegraphics[clip=true,width=8cm]
{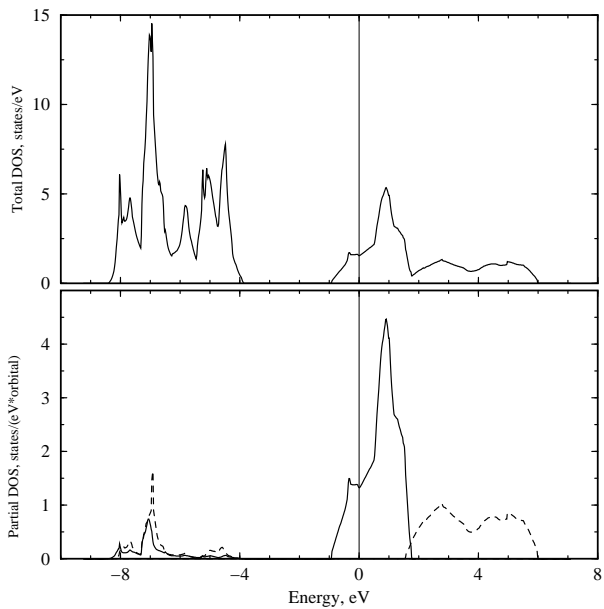}

 \caption{Densities of states of LaTiO$_3$ calculated with LDA-LMTO.
Upper
figure: total DOS; lower figure: partial $t_{2g}$ (solid lines)
and $e_g$ (dashed lines) DOS [reproduced from Ref.\protect
\cite{Nekrasov00}].} \label{ldados}
\end{figure}

\begin{figure}[tbp]
\centering
\includegraphics[clip=true,width=8cm]
{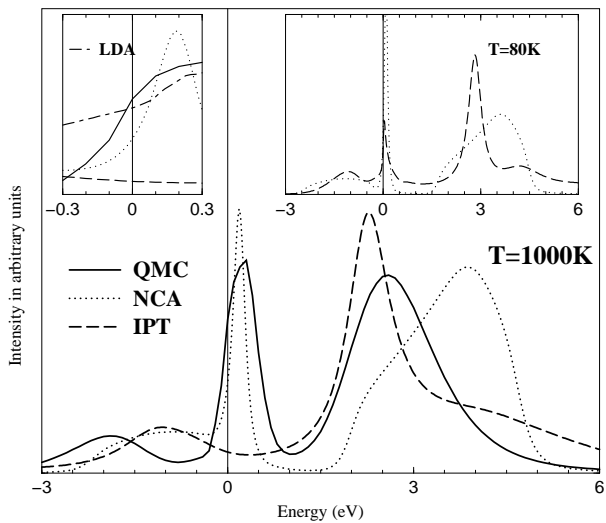}

\caption{Spectrum  of La$_{0.94}$Sr$_{0.06}$TiO$_{3}$
 as calculated by LDA+DMFT(X) at $T=0.1$~eV ($\approx 1000$~K)
and $U=4$~eV employing the approximations X=IPT, NCA, and
numerically exact QMC. Inset left: Behavior at the Fermi level
including the LDA DOS. Inset right:  X=IPT and NCA spectra at
$T=80$~K
 [reproduced from Ref.\protect \cite{Nekrasov00}].}
\label{dmft_latio}
\end{figure}

The LDA band-structure calculation for undoped (cubic) LaTiO$_{3}$
yields the DOS shown in Fig. \ref{ldados} which is typical for
early transition metals. The oxygen bands, ranging from $-8.2$~eV
to $-4.0$~eV, are filled such that Ti is three-valent. Due to the
crystal-field splitting, the Ti 3$d$-bands separates into two
empty $e_{g}$-bands and three degenerate $t_{2g}$-bands. Since the
$t_{2g}$-bands at the Fermi energy are well separated also from the
other bands we employ the approximation introduced in section 2.5
which allows us to work with the LDA DOS [Eq.~(\ref{intg})]
instead of the full one-particle Hamiltonian [Eq.~(\ref{Dyson})].
In the LDA+DMFT calculation, Sr-doping $x$ is taken into account
by adjusting the
chemical potential to yield $n=1-x=0.94$ electrons within the $t_{2g}$%
-bands.  
There is some uncertainty in the LDA-calculated Coulomb
interaction parameter $U$ $\sim 4-5$ eV (for a discussion see Ref.\cite%
{Nekrasov00}) which is here assumed to be spin- and
orbital-independent.
In Fig.~\ref{dmft_latio}, results for the spectrum of
La$_{0.94}$Sr$_{0.06}$%
TiO$_{3}$ as calculated by LDA+DMFT(IPT, NCA, QMC) for the same LDA DOS
at $%
T\approx 1000$~K and $U=4$~eV are compared\cite{Nekrasov00}. In
Ref.\cite{Nekrasov00} the formerly presented IPT\cite{poter97} and
NCA\cite{zoelfl99} spectra were recalculated to allow for a
comparison at exactly the same parameters. All three
methods yield the
typical features of strongly correlated metallic paramagnets: a
lower Hubbard band, a quasi-particle peak (note that IPT produces
a quasi-particle peak only below about 250K which is therefore not
seen here), and an upper Hubbard band. By contrast, within LDA the
correlation-induced Hubbard bands are missing and only a broad
central quasi-particle band (actually a one-particle peak) is
obtained (Fig.~\ref{ldados}).

While the results of the three evaluation techniques of the DMFT
equations (the approximations IPT, NCA and the numerically exact
method QMC) agree on a qualitative level, Fig.~\ref{dmft_latio}
reveals considerable quantitative differences. In particular, the
IPT quasi-particle peak found at low temperatures (see right inset
of Fig.~\ref{dmft_latio}) is too narrow such
that it disappears already at about 250~K and is, thus, not present at $%
T\approx 1000$~K. A similarly narrow IPT quasi-particle peak was
found in a
three-band model study with Bethe-DOS by Kajueter and Kotliar\cite{Kajueter}%
. Besides underestimating the Kondo temperature, IPT also produces
notable deviations in the shape of the upper Hubbard band.
Although NCA comes off much better than IPT it still
underestimates the width of the quasiparticle peak by a factor of
two. Furthermore, the position of the quasi-particle
peak is too close to the lower Hubbard band. In the left inset of Fig.~\ref%
{dmft_latio}, the spectra at the Fermi level are shown. At the
Fermi level, where at sufficiently low temperatures the
interacting DOS should be pinned at the non-interacting value, the
NCA yields a spectral function which is almost by a factor two too
small. The shortcomings of the NCA-results appear to result from
the well-known problems which this approximation scheme encounters
already in the single-impurity Anderson model at low temperatures
and/or low frequencies\cite{MH,NCAdeficit}. Similarly, the
deficiencies of the IPT-results are not entirely surprising in
view of the semi-phenomenological nature of this approximation,
especially for a system off half filling. This comparison shows
that the choice of the method used to solve the DMFT equations is
indeed important, and that, at least for the present system, 
the approximations IPT and NCA differ quantitatively from the
numerically exact QMC.

\begin{figure}[tbp]
 \centering
\includegraphics[clip=true,width=8cm]
 {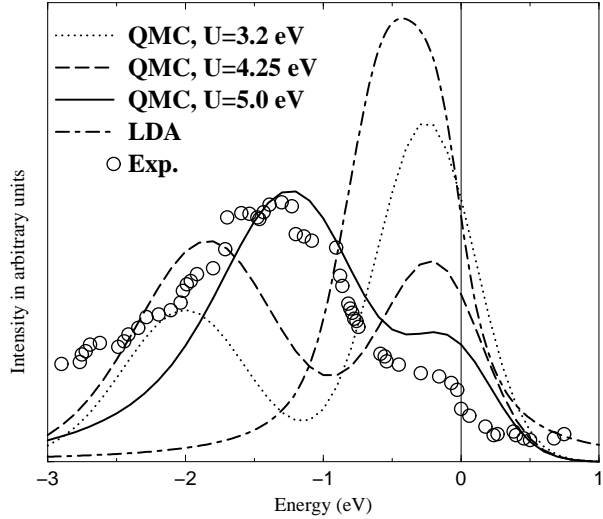}

\caption{Comparison of the experimental photoemission
spectrum\protect\cite{fujimori}, the  LDA result, and the
LDA+DMFT(QMC) calculation for
 La$_{0.94}$Sr$_{0.06}$TiO$_{3}$ (i.e., 6\% hole doping) and different
Coulomb interaction $U=3.2$, $4.25$, and $5$~eV [reproduced from
Ref.\protect\cite{Nekrasov00}]. \label{explatio}}
\end{figure}

Photoemission spectra provide a direct experimental tool to study
the electronic structure and spectral properties of electronically
correlated materials. A comparison of LDA+DMFT(QMC) at
1000~K\cite{Note} with the
experimental photoemission spectrum\cite{fujimori} of La$_{0.94}$Sr$_{0.06}$%
TiO$_{3}$ is presented in Fig~\ref{explatio}. To take into account
the uncertainty in $U$\cite{Nekrasov00}, we present results for
$U=3.2$, $4.25$ and $5$ eV. All spectra are multiplied with the
Fermi step function and are Gauss-broadened with a broadening
parameter of 0.3~eV to simulate the experimental
resolution\cite{fujimori}. LDA band structure calculations, the
results of which are also presented in Fig.~\ref{explatio},
clearly fail to reproduce the broad band observed in the
experiment at 1-2~eV below the Fermi energy\cite{fujimori}. Taking
the correlations between the electrons into account, this lower
band is easily identified as the lower Hubbard band whose spectral
weight originates from the quasi-particle band at the Fermi energy
and which increases with $U$. The best agreement with experiment
concerning the relative intensities of the Hubbard band and the
quasi-particle peak and, also, the position of the Hubbard band is
found for $U=5$ eV. The value $U=5$ eV is still compatible with
the {\em ab initio} calculation of this parameter within
LDA\cite{Nekrasov00}. One should also bear in mind that
photoemission experiments are sensitive to surface properties. Due
to the reduced coordination number at the surface the bandwidth is
likely to be smaller, and the Coulomb interaction less screened,
i.e., larger. Both effects make the system more correlated and,
thus, might also explain why better agreement is found for $U=5$
eV. Besides that, also the polycrystalline nature of the sample,
as well as spin and orbital\cite{Keimer} fluctuation not taken
into account in the LDA+DMFT approach, will lead to a further
reduction of the quasi-particle weight.

\pagebreak

\section{Conclusion}

In this paper we described the set-up and presented results of the
computational scheme LDA+DMFT which merges two non-perturbative,
complementary investigation techniques for many-particle systems
in condensed matter physics. This approach allows one to perform
{\em ab initio }calculations of strongly correlated electronic
systems. Using the band structure results calculated within local
density approximation (LDA) as input, the missing electronic
correlations are introduced by dynamical mean-field theory (DMFT).
On a technical level this requires the solution of an effective
self-consistent, multi-band Anderson impurity problem by some
numerical method (e.g. IPT, NCA, QMC). Comparison of 
 the photoemission spectrum of La$_{1-x}$Sr$_{x}$TiO$_{3}$
calculated by LDA+DMFT using IPT, NCA and QMC reveal that the
choice of the evaluation method is of considerable importance.
Indeed, only with the numerically exact QMC quantitatively
reliable results are obtained. The results of the LDA+DMFT(QMC)
approach are found to be in very good agreement
with the experimental photoemission spectrum of La$_{0.94}$Sr$_{0.06}$TiO$%
_{3}$.

The LDA+DMFT(QMC) scheme will provide a powerful tool for all
future {\em ab initio} investigations of real materials with
strong electronic correlations. In particular, LDA+DMFT is the
only existing {\em ab initio} approach which is able to
investigate correlated electronic systems close to a Mott
transition, as well as heavy fermion and $f$-electron materials.
The physical properties of such systems are characterized by the
correlation-induced generation of small, Kondo-like energy scales
which require the application of genuine many-body techniques.

With LDA+DMFT the band structure and model Hamiltonian communities
which essentially lived separate lives so far are finally able to
join forces. In the future both communities will need each other's
input. Indeed, without DMFT the LDA will be limited to the
investigation of weakly correlated systems, and without the
LDA-calculated band structure the many-body approach will be
restricted to the study of more or less oversimplified models.


\nonumsection{Acknowledgment}
\noindent
We are grateful to R. Claessen, V. Eyert,
J. L\ae gsgaard, A. Lichtenstein, A.K. McMahan,
Th.~Pruschke, G.A. Sawatzky, J. Schmalian, and
M. Z\"olfl for useful discussions.
This work was supported in part by the Sonderforschungsbereich
484 of the Deutsche Forschungsgemeinschaft,
the Russian Foundation for Basic Research (RFFI-98-02-17275),
and a Feodor-Lynen grant of the AvH foundation.

\nonumsection{References}


\begin{thebibliography}{99}
\bibitem{nonlocalU} Note, that the largest non-local contribution is the
nearest-neighbor density-density interaction which, to leading order in
$Z$ ($Z$: number of nearest-neighbor sites), yields only the Hartree
term\cite{MH1} which is already taken into account in the LDA.
\bibitem{QMCdb}
We note that QMC
is restricted to temperatures of the order of $T=0.1$~eV ($\approx
$~1000~K) on current workstations. Furthermore, it relies on the
maximum entropy method \cite{MEM} for the analytical continuation
from Matsubara to real frequencies. This method is reliable in the
vicinity of the Fermi energy whereas its energy resolution becomes
poorer at larger energies.
\bibitem{NCAdeficit}  NCA is a self-consistent perturbation theory for the
single-impurity Anderson model and is often a fast method to obtain results
for this model and also within DMFT. However, NCA is known to violate Fermi
liquid properties at temperatures much lower than the smallest energy scale
of the problem and whenever charge excitations become important\cite{MH,Pruschke93}.
Hence, in some parameter ranges it fails to account for the
proper low-energy physics and must therefore be applied with considerable
care\cite{Pruschke93}.
\bibitem{Note}
At present, QMC simulations of the DMFT equations
 are not feasible at the experimental temperature
(80K). We note, however, that no
intrinsic temperature dependence was observed in the experiment\cite{fujimori},
at least up to room temperature.
\end{thebibliography}
\end{document}